\documentclass[preprintnumbers,amsmath,nofootinbib,amssymb]{revtex4}
\usepackage{color}
\usepackage{graphicx}
\usepackage{dcolumn}
\usepackage{bm}

\begin{document}



\begin{center}
{\Large {\bf Scalar Dark Matter Effects in Higgs and Top Quark
Decays}}
\end{center}

\begin{center}
{Xiao-Gang He$^{1}$, Tong Li$^2$, Xue-Qian Li$^2$, and Ho-Chin
Tsai$^1$}\\
\vspace*{0.3cm}

$^1$Department of Physics and Center for Theoretical Sciences,
National Taiwan University, Taipei\\

$^2$Department of Physics, Nankai University, Tianjin, 300071
\footnote{hexg@phys.ntu.edu.tw; allongde@mail.nankai.edu.cn;
lixq@nankai.edu.cn; hctsai@phys.ntu.edu.tw}
\end{center}


\begin{center}
\begin{minipage}{12cm}

\noindent Abstract:\\
We study possible observational effects of scalar dark matter, the
darkon $D$, in Higgs $h$ and top quark $t$ decay processes, $h\to
DD$ and $t\to c DD$ in the minimal Standard Model (SM) and its two
Higgs doublet model (THDM) extension supplemented with a SM
singlet darkon scalar field $D$. We find that the darkon $D$ can
have a mass in the range of sub-GeV to several tens of GeV,
interesting for LHC and ILC colliders, to produce the required
dark matter relic density.  In the SM with a darkon, $t \to c DD$
only occurs at loop level giving a very small rate, while the rate
for Higgs decay $h\to DD$ can be large. In THDM III with a darkon,
where tree level flavor changing neutral current (FCNC)
interaction exists, a sizable rate for $t\to c DD$ is also
possible.

\end{minipage}

\end{center}

\section{Introduction}

Understanding the nature of dark matter is one of the most
challenging problems in particle physics and cosmology. Although
dark matter contributes about 20\% to the energy density of our
universe\cite{cosm}, the identity of the basic constituents of the
dark matter is still not known. One of the popular candidates for
dark matter is the Weakly Interacting Massive Particle (WIMP).
Detection of WIMP candidate is extremely important in
understanding the nature of dark matter and also the fundamental
particle physics models. The traditional way is to measure the
dark matter flux at earth detectors. It is interesting to see
whether WIMP can be produced and detected at collider experiments
directly.  Among the many possible WIMPs, the lightest
supersymmetric particle is the most studied one. But no direct
experimental evidence has been obtained for supersymmetry so that
other possibilities of WIMP which explain dark matter relic
density in our universe should be studied and searched for.

The simplest model which has a candidate of WIMP is the Standard
Model (SM) with a singlet SM real scalar field $D$ (SM+D). We will
call the field D as darkon. The darkon field as dark matter was
first considered by Silveira and Zee\cite{Zee1}, and further
studied later by several others
groups\cite{Zee2,McD,bento,Posp1,Posp,Zhu}. In this work we
concentrate on the darkon observable effects on Higgs $h$ and top
quark $t$ decays. We find that a darkon of mass in the range of
sub GeV to tens of GeV can play the role of dark matter with very
constrained parameters and also have interesting collider physics
signatures. We will also extend the studies to two Higgs doublet
models (THDM). The LHC to be in operation soon and the planed ILC
offer excellent possibilities to study darkon signatures through
$h\to DD$ and $t\to c DD$.

\section{Dark Matter Constraints on Darkon}

The darkon field $D$ must interact weakly with the standard matter
field sector to play the role of dark matter. The simplest way of
introducing the darkon $D$ is to make it a SM real singlet which
can only be created and annihilated in pairs, the SM+D model. If
the interaction of $D$ is required to be renormalizable, it can
only couple to the Higgs doublet field $H$. Beside the kinetic
energy term $-(1/ 2)\partial_\mu D\partial^\mu D$, the general
form of other terms are given by\cite{Posp1,Posp}
\begin{eqnarray}
-\mathcal{L}_D &=& {\lambda_D\over 4}D^4+{m_0^2\over 2}D^2+\lambda
D^2H^{\dag}H\;.\label{DH}
\end{eqnarray}

Note that the above Lagrangian is invariant under a $D \to - D$
$Z_2$ symmetry. The parameters in the potential should be chosen
such that the $D$ field will not develop vacuum expectation value
(vev) and the $Z_2$ symmetry is not broken, after $SU(2)_L\times
U(1)$ spontaneously breaks down to $U(1)_{em}$, to make sure that
darkons can only be produced or annihilated in pairs.
The relic density of $D$ is
then decided, to the leading order, by annihilation of a pair of
$DD$ into SM particles through Higgs
exchange\cite{Zee2,Posp1,Zhu}, $DD \to h \to X$ where $X$
indicates SM particles.

Eliminating the pseudo-goldstone boson ``eaten'' by $W$ and $Z$,
we have the physical Higgs h coupling to $D$ as
\begin{eqnarray}
 -L_D&=&{\lambda_D\over
4}D^4+{1\over 2}( m_0^2 + \lambda v^2) D^2 +{1\over 2}\lambda h^2
D^2 + \lambda v h D^2,
\end{eqnarray}
where $v=246$ GeV is the vev of $H$. The $D$ field has a mass
$m^2_D = m^2_0 + \lambda v^2$. The last term $\lambda v h D^2$
plays an important role in determining the relic density of the
dark matter.

The annihilation of a $DD$ pair into SM particles is through
s-channel h exchange. To have some idea how this works, let us
consider $DD \to h \to f\bar f$. We parameterize Higgs-fermion and
Higgs-darkon interactions as
\begin{eqnarray}
- L_Y = a_{ij} \bar f_R^i f_L^j h + b h D^2\;,\label{coupling-f}
\end{eqnarray}
where $R(L) = (1\pm\gamma_5)/2$. In the SM, $a_{ij} =
m_i\delta_{ij}/v$ and $b = \lambda v$.

The total averaging annihilation rate of a pair $DD$ to fermion
pairs is then given by
\begin{eqnarray}
\langle v_r \sigma\rangle  = {16 b^2 \over 32 \pi m^3_D} {1\over
(4 m^2_D -m^2_h)^2 + \Gamma^2_h m^2_h} \sum_f N^c_f |a_{ff}|^2 (4
m_D^2 - 4 m^2_f)^{3/2}.
\end{eqnarray}
where $N^c_f$ is the number of colors of the f-fermion. For a
quark $N^c_f = 3$ and a lepton $N^c_f = 1$. $f$ sums over the
fermions with $m_f < m_D$. In the above $v_r$ is the average
relative velocity of the two $D$ particles. We have used the fact
that for cold dark matter D, the velocity is small, therefore to a
good approximation the average relative speed of the two $D$ is
$v_r = 2 p_{_{Dcm}}/m_D$ and $s = (p_{f} + p_{\bar f})^2$ is equal
to $4m_D^2$ .

If there are other decay channels, the sum should also include
these final states. The above can be re-written and generalized
to\cite{Posp}
\begin{eqnarray}
\langle v_r \sigma \rangle = {8 b^2\over (4m^2_D -m_h^2)^2+m^2_h
\Gamma^2_h} {\Gamma(\tilde h \to X')\over 2 m_D},
\end{eqnarray}
where $\Gamma(\tilde h\to X') =\sum_i \Gamma(\tilde h \to X_i)$
with $\tilde h$ being a ``virtual'' Higgs having the same
couplings to other states as the Higgs $h$, but with a mass of
$2m_D$. $X_i$ indicate any possible decay modes of $\tilde h$. For
a given model $\Gamma(\tilde h \to X')$ is obtained by calculating
the $h$ width and then set the mass equal to $2 m_D$.

To produce the right relic density for dark matter $\Omega_D$, the
annihilate rate needs to satisfy the following\cite{turner}
\begin{eqnarray}
&&\langle v_r \sigma \rangle \approx {1.07\times 10^9 x_f\over
\sqrt{g_*} m_{pl} \mbox{GeV} (\Omega_D h^2)}\;,\;\;x_f \approx
\ln{0.038 m_{pl} m_D \langle v_r \sigma \rangle \over \sqrt{g_*}
x_f},
\end{eqnarray}
where $m_{pl} = 1.22 \times 10^{19}$ GeV, $x_f = m_D/T_f$ with
$T_f$ being the freezing temperature, and $g_*$ is the
relativistic degrees of freedom with mass less than $T_f$. Note
that the `h' in $\Omega_D h^2$ is the normalized Hubble constant,
not the Higgs field.

For given values of $m_D$ and $\Omega_D h^2$, $x_f$ and $g_*$ can
be determined and therefore also $\langle v_r \sigma \rangle$.
Then one can determine the parameter $b$. In Fig. 1 we show the
allowed range for the parameter $b/v=\lambda$ as a function of the
darkon mass $m_D$ for several values of Higgs mass $m_h$ with
$\Omega_Dh^2$ set in the range $0.095 \sim 0.112$ determined from
cosmological observations\cite{cosm}. We see that the darkon mass
can be as low as a GeV.  Since we are interested in producing the
darkons and study their properties at colliders, we will limit
ourselves to study darkon with a mass less than 100 GeV. We note
that when the darkon mass decreases, $\lambda$ becomes larger. For
small enough $m_D$ $\lambda$ can be close to one which may upset
applicability of perturbative calculation. We will only show
region of parameters with $\lambda <1$.


\begin{figure}[!htb]
\begin{center}
\begin{tabular}{cc}
\includegraphics[width=2.7in]{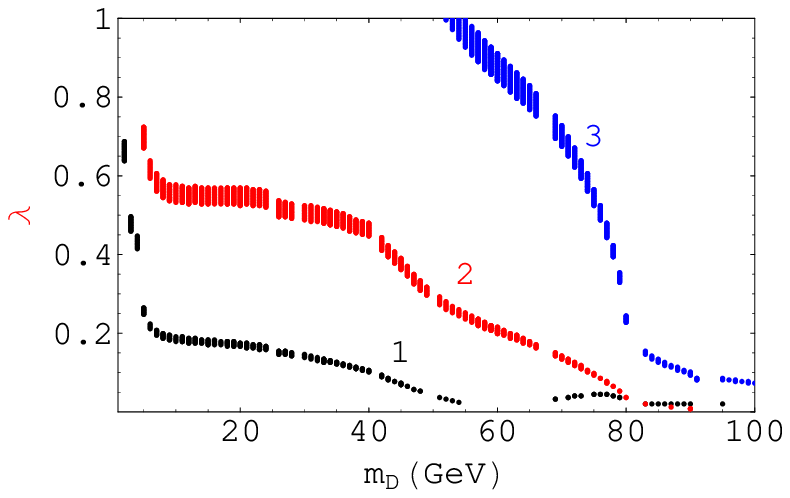}
\includegraphics[width=2.7in]{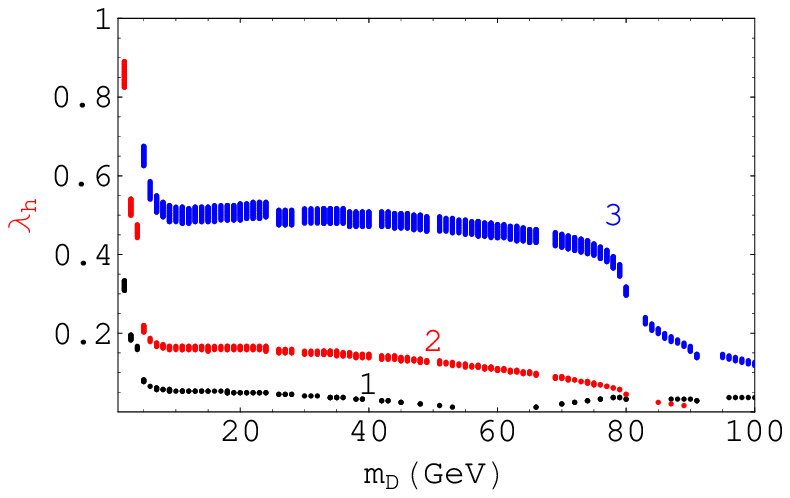}
\end{tabular}
\end{center}
\caption{ The $\lambda_h$ in SM+D model (left) and THDM III+D
model (right) as a function of $m_D$ where the shelded areas 1
(black), 2 (red), 3 (blue) are for $m_h =$120, 200 and 350GeV,
respectively (same for other figures). For THDM III+D, we have
assumed the physical Higgs h to be the lightest one and
$tan\alpha=1, tan\beta=5$ GeV are used in
2HDM III+D model for illustration. \protect\label{bv}}
\end{figure}

\section{Effects of Darkon on Higgs and Top Decays}

At LHC and ILC, a large number of Higgs and top quark particles
may be produced if kinematically accessible\cite{EXP,SMhiggs}. The
various production cross sections of Higgs at LHC and ILC are
typically a few pb level\cite{SMhiggs}. Assuming the integrated
luminosities at LHC and ILC to be $200 fb^{-1}$, a large number of
Higgs can be copiously produced and its properties studied in
details. The main effect of the darkon field on the Higgs
properties is to add an invisible decay mode $h\to DD$ to the
Higgs particle. Due to this additional mode, the Higgs width will
be broader affecting determinations of the Higgs mass, and also
decay properties in processes such as $pp\to X h\to X X'$
and\cite{Posp1} $e^+ e^- \to Z^*\to Z h \to Z X'$. Here $X'$
indicates the final states used to study $h$ properties.

Top quark properties will be studied in details at LHC and ILC.
Assuming an integrated luminosity of $200fb^{-1}$, the sensitivity
for $B(t \to ch)$ can reach $3 \times 10^{-5}$ at
LHC\cite{EXP,tchLHC} and  $4.5 \times 10^{-5}$ at
ILC\cite{EXP,tchLHC}. Branching ratio for $t\to c DD$ at that
level will therefore significantly affect the results and should
be accounted. The decay mode $t\to c DD$ is interesting for
several reasons. Since $D$ is neutral if it is produced without
tagging it is not possible to identify its production. Through top
decay, one can in some way to tag it if one considers pair
production of top\cite{hegunion}. The signal will be a charm jet
plus missing energy. There may be other processes producing
similar signal, such as $t\to c \bar \nu \nu$. This decay mode in
the SM is, however, very small\cite{neutrino}. Another reason for
studying this flavor changing neutral current (FCNC) process is
that the top quark is heavy, the mass of D up to about 80 GeV can
be studied compared with other quark decays, such as b quark decay
where mass of D below 2.5 GeV may be studied \cite{Posp}.

In our previous discussions on dark matter density we have seen
that the coupling $\lambda = b/v$ in a wide range of darkon mass
is not much smaller than 1, it is clear that the introduction of
darkon will affect processes mediated by Higgs exchange and Higgs
decay itself. In Figs. 2 and 3 we show the decay width
$\Gamma(h\to DD)$ and the branching ratio $B(h\to DD)$ as a
function of $m_D$ for several values of $m_h$. We see that the
invisible decay $h\to DD$ dominates over the Higgs decay width if
$m_D$ is significantly below the $h \to DD$ threshold. However such
invisible domination becomes weaker when $h \to VV$ modes become
kinematically allowed. This will affect the bounds set on the
Higgs mass for low mass $m_D$.


\begin{figure}[!htb]
\begin{center}
\begin{tabular}{cc}
\includegraphics[width=2.7in]{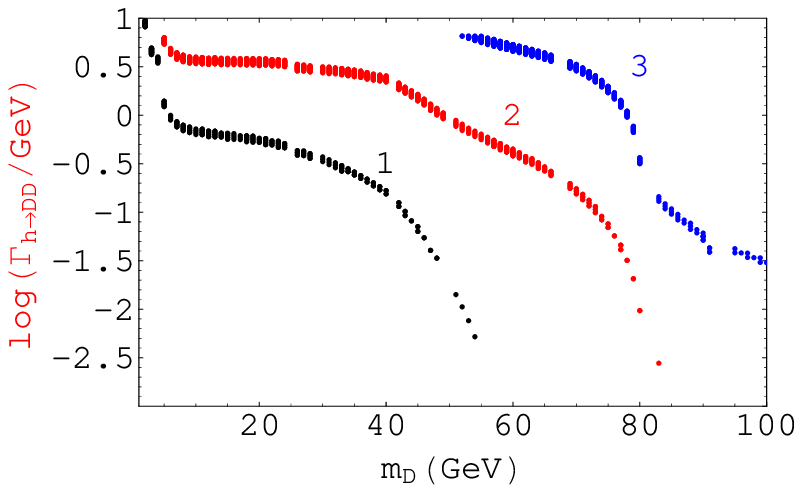}
\includegraphics[width=2.7in]{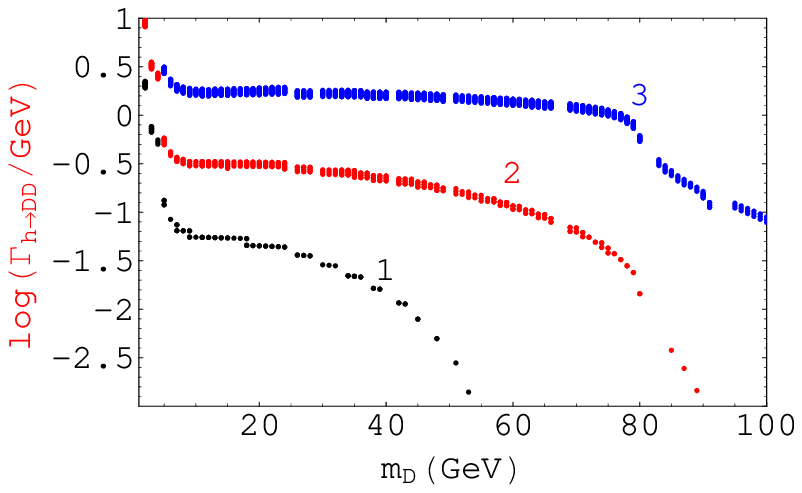}
\end{tabular}
\end{center}
\label{figure} \caption{ The decay widths of $h \to DD$ in SM+D
(left) and THDM III+D (right) as a function of $m_D$.
\protect\label{hDD}}
\end{figure}


\begin{figure}[!htb]
\begin{center}
\begin{tabular}{cc}
\includegraphics[width=2.7in]{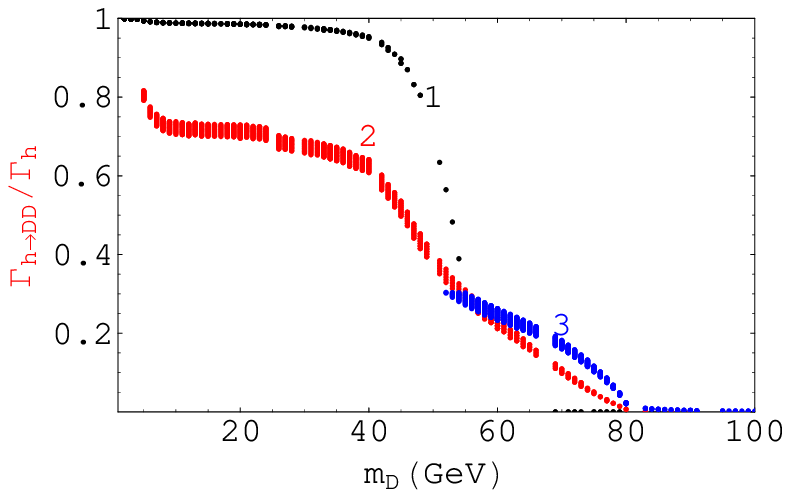}
\includegraphics[width=2.7in]{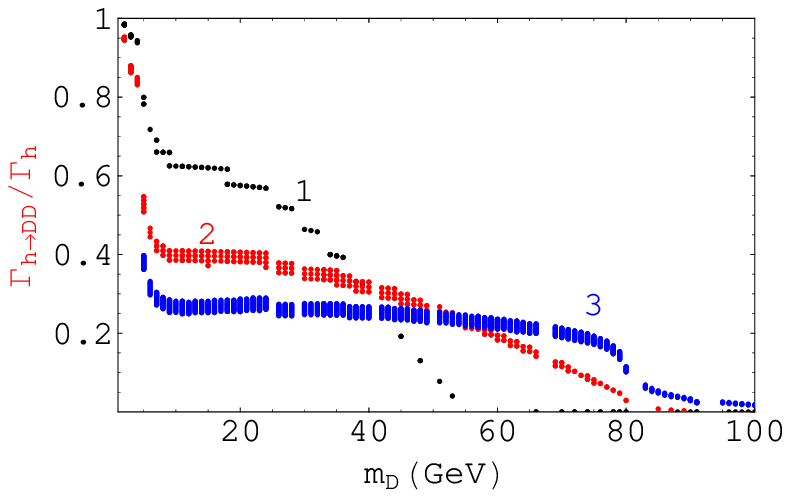}
\end{tabular}
\end{center}
\label{figure} \caption{The
 branching ratios of $h \to DD$ in SM+D (left) and THDM III+D (right) as
a function of $m_D$. \protect\label{BRhDD}}
\end{figure}

For Higgs mediated $t\to c DD$, one needs to know the couplings
$a_{ij}$ defined in eq.(\ref{coupling-f}). The decay amplitude is
\begin{eqnarray}
&&M(f_i \to f_j DD)  = { 2 b\over s-m^2_h + i \Gamma_h m_h} \bar
f_j ( a_{ji} R + a^*_{ij} L ) f_i.
\end{eqnarray}

In the SM+D, flavor changing coupling of Higgs to fermions are
generated at loop level and therefore small. Using the expression
in Ref. \cite{Soni} for the SM Higgs couplings to fermions, we
find that the branching ratio $B(t\to c h \to c DD)$ to be less
than $10^{-13}$ for the Higgs h mass in the hundred GeV range.
This is too small to be observed by future experiments at either
LHC or ILC.

To have detectable effects for $t\to c DD$, there must be new
physics beyond the SM+D model where tree level FCNC interaction
mediated by Higgs exists. To this end we take the two Higgs
doublet model as an example for discussion in the following.

\section{Two Higgs Doublet Models with Darkon}

Depending on how the two Higgs doublets $H_{1,2}^T = (h^+_{1,2},
(v_{1,2} + h_{1,2} + i a_{1,2})/\sqrt{2}) $ couple to quarks and
leptons, there are different models with darkon field added
(THDM+D). We will come back to this later. We first discuss how
darkon $D$ couples to the two Higgs doublets. In analog to
(\ref{DH}), we write down the most general renormalizable
interaction of $D$ with the Higgs doublet fields $H_{1,2}$ in the
Higgs potential. We have
\begin{eqnarray}
-\mathcal{L}_D &=& {\lambda_D\over 4}D^4+{m_0^2\over 2}D^2+
D^2(\lambda_1 H_1^{\dag}H_1 + \lambda_2 H^\dagger_2 H_2 +
\lambda_3 (H^\dagger_1 H_2 + H^\dagger_2 H_1)).
\end{eqnarray}
We have again imposed the $Z_2$ symmetry previously discussed. For
the same reason we need to keep it unbroken. We have also assumed
CP conservation in the above Lagrangian.

Eliminating the pseudo-goldstone boson ``eaten'' by $W$ and $Z$,
the two Higgs doublets have physical components\cite{2hdm}, $H_1^T
= (
 -\sin\beta h^+, (v_1+ \cos\alpha H - \sin\alpha h - i \sin\beta A)/\sqrt{2})$, and $
H_2^T = ( \cos\beta h^+, (v_2 + \sin\alpha H + \cos \alpha h + i
\cos \beta A)/\sqrt{2} )$, where $\tan\beta = v_2/v_1$. $\alpha$
is the mixing angle for the scalar Higgs fields $h$ and $H$ with
$h$ playing a similar role as the SM Higgs. $A$ is a physical
pseudoscalar field.

Using the above information, we obtain the mass of $D$ and the $h
D^2$ interaction after $H_{1,2}$ develop vevs,
\begin{eqnarray}
&&m^2_D = m^2_0 + v^2 (\lambda_1 \cos^2\beta +\lambda_2
\sin^2\beta
+2\lambda_3 \cos\beta \sin\beta),\nonumber\\
&&- L_{hD^2} = [-\lambda_1 \cos\beta \sin\alpha + \lambda_2
\sin\beta \cos\alpha + \lambda_3 \cos(\beta +\alpha)]v h D^2
=\lambda_h v h D^2.
\end{eqnarray}
The mass parameter $m_D^2$ and the effective coupling $\lambda_h$
are free parameters in this model.

The couplings of $H$ and $A$ to $DD$ are: $-L_{HD^2} = (\lambda_1
\cos\beta \cos\alpha + \lambda_2 \sin\beta \sin\alpha + \lambda_3
\sin(\beta +\alpha))v H D^2=\lambda_H v H D^2$, and $-L_{AD^2}=0$.
If both the two neutral scalar particles $h$ and $H$ contribute,
the analysis will be complicated. For concreteness, in our
numerical analysis we will neglect contributions from the $H$ by
requiring small $\lambda_H$. In this case as far as dark matter
relic density is concerned, one can treat $\lambda_h$ as an
effective coupling $\lambda_h = b/v$ defined before. The
constraint on $\lambda_h$ can be obtained in a similar way as that
for SM+D. The results for illustrating parameters chosen are shown
in Fig.1. We note that $\lambda_h$ in this case needs not to be as
large as that in the SM+D model because, for the parameters used,
the width of Higgs decay to SM particle for small D mass is
larger.

The couplings of the Higgs fields to quarks distinguish different
models\cite{2hdm}. In the literature, these different models are
called THDM I, THDM II and THDM III which are defined by only one
Higgs gives masses to both up and down quark masses, $H_1$ gives
down quark and $H_2$ gives up quark masses, and both $H_1$ and
$H_2$ give up quark and down quark masses, respectively. In models
I and II $t\to c D D$ are generated at one loop level hence the
decay rate are too small to be detected at LHC and ILC\cite{tchLHC,2hdm},
although can be substantially larger than that predicted by SM.
THDM III offers a possibility to have a large detectable rate. We
therefore will only show some details for THDM III. We will refer
this model as THDM III+D. The couplings of h to fermions are given
by\cite{2hdm}
\begin{eqnarray}
L_{III} &=& -\bar Q_L \lambda^u_1 \tilde H_1 U_R - \bar Q_L
\lambda^u_2 \tilde H_2 U_R -\bar Q_L \lambda^d_1  H_1 D_R - \bar
Q_L \lambda^d_2 H_2 D_R \nonumber\\
&& -\bar L_L \lambda^l_1  H_1 E_R - \bar L_L \lambda^l_2 H_2 E_R+
h.c.\;,
\end{eqnarray}
where $\tilde H_i = i \tau_2 H^*_i$.

We obtain the $h$ coupling to fermions as
\begin{eqnarray}
L_{III} &=& -\bar U_{L} M^uU_{R} {\cos \alpha \over v \sin \beta}h
+ \bar U_{L}\tilde  M^u U_{R} {\cos(\alpha -\beta)\over v \sin
\beta }
 h+  \bar D_{L} M^d D_{R}{\sin \alpha\over v \cos \beta} h \\
 & -& \bar
D_{L} \tilde M^d D_{R} { \cos(\alpha -\beta)\over v \cos \beta }
 h + \bar E_{L} M^l E_{R} {\sin \alpha\over v \cos \beta} h - \bar
E_{L} \tilde M E_{R} { \cos(\alpha -\beta)\over v \cos \beta } h +
h.c. \;,\nonumber
\end{eqnarray}
where $M^{u,d,l}=(\lambda_1^{u,d,l}v_1 +
\lambda_2^{u,d,l}v_2)/\sqrt{2}$ are the diagonalized masses of the
up-quarks, down-quarks and charged leptons. The off diagonal
entries $\tilde M^u = \lambda^u_1v/\sqrt{2}$ and $\tilde M^{d,l}
=\lambda^{d,l}_2 v/\sqrt{2}$ are not fixed. We have chosen a
parametrization for $\tilde M^i$ in the limit that when they are
set to zero, the Yukawa couplings reduce to the minimal SUSY ones.
In our later discussions, we will follow Ref.\cite{jaas} to
parameterize the off diagonal entries to have the geometric mean
form $\tilde M_{ij}^{u,d,l}=\rho_{ij}^{u,d,l}{\sqrt{m_im_j}}$ with
$\rho_{ij}\simeq1$ for concreteness, and $\rho_{ii}$ to be
negligibly small, for illustration.

The couplings of $h$ to $W$, $Z$ will also be changed to
\begin{eqnarray}
L_{hWW} = {2m_W^2\over v} sin(\beta-\alpha) h W^2\;,\;\; L_{hZZ} =
{m_Z^2\over v} sin(\beta-\alpha) h Z^2\;,
\end{eqnarray}
which affect the Higgs decay width.

Using the above information, we can obtain the total $h$ decay
width without and with the $h \to DD$ mode. The resulting
$\Gamma(h \to DD)$ and $B(h\to DD)$ are shown in Figs. 2 and 3. We
see that $\Gamma(h \to DD)$ and $B(h\to DD)$ in THDM III+D model
can be smaller than those in the SM+D because $\Gamma(\tilde h \to
X')$ in Eq.(5) is bigger in THDM III+D when $m_D$ is small.

A large difference of THDM III+D compared with SM+D can show up in
$t \to cDD$ decays. The results are shown in Fig. 4. For the above
parametrization of $h$ coupling to fermions, we find that the
branching ratio $B(t\to c DD)$ can be as large as $10^{-3}$ if $h$
mass is below the $h \to VV$ threshold which can be investigated
at LHC and ILC\cite{tchLHC}.


\begin{figure}[!htb]
\begin{center}
\begin{tabular}{cc}
\includegraphics[width=2.7in]{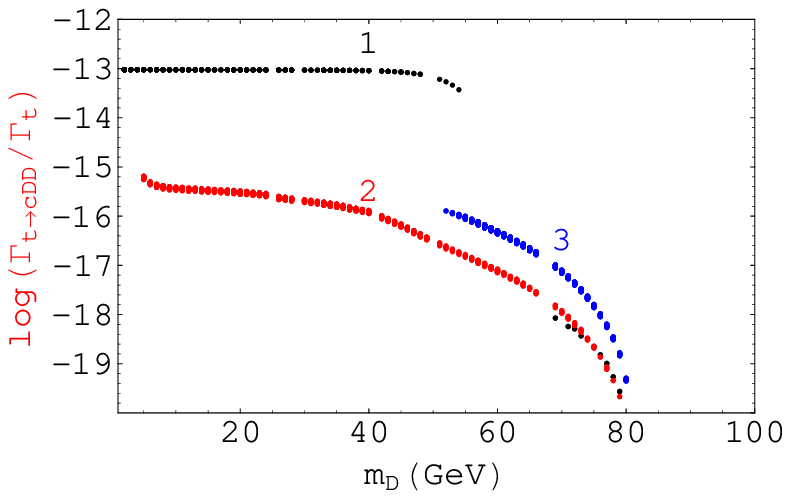}
\includegraphics[width=2.7in]{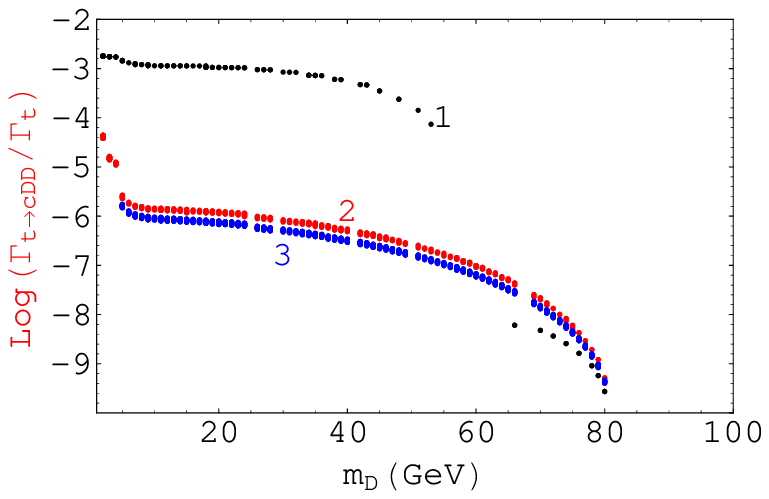}
\end{tabular}
\end{center}
\label{figure} \caption{The branching ratios of $t\to cDD$ in SM+D
(left) and THDM III+D (right) as a function of $m_D$.
\protect\label{BRtcDDIII}}
\end{figure}

\section{Conclusions}

We have studied the effects of scalar dark matter, the darkon $D$,
in Higgs $h$ and top quark $t$ decay processes, $h\to DD$ and
$t\to c DD$ in the SM+D and THDM III+D. Requiring renormalizable
interaction for these models, the darkon field can only couple to
the Higgs field in the Lagrangian. We find that the darkon $D$ can
have a mass in the range of sub-GeV to several tens of GeV,
interesting for LHC and ILC colliders, to produce the required
dark matter relic density with restricted darkon and Higgs
coupling.

In both SM+D and THDM III+D models, the darkon field can have
significant effects on the Higgs decay width through the invisible
Higgs decay mode $h \to DD$. In the SM+D, if the darkon mass is
significantly lower than the $h \to DD$ threshold, this invisible
decay width dominates the Higgs decay making determination of
Higgs mass in this region more difficult than that in the SM. In
the THDM III+D, the invisible decay width can be made not so
dominating.

In the SM+D model, $t \to c DD$ only occurs at loop level giving a
very small rate. In THDM III+D, where tree level FCNC interaction
exists, a sizable rate of order $10^{-3}$ for $t\to c DD$ is
possible for $m_h$ below $2 m_W$. Experiments at LHC and ILC will
be able to provide important information about darkon field
interaction.

\section*{Acknowledgments}
This work was partially supported by NSC, NCTS and NNSF. XGH
thanks A. Zee for many useful discussions and also suggesting the
name `Darkon' for the scalar dark matter.


\begin{thebibliography}{0}
\bibitem{cosm}D.N. Spergel {\it et al.} [arXiv: astro-ph/0603449];
W.-M. Yao, {\it et al.}, Particle Data Group, J. Phys. {\bf G33},
1 (2006).

\bibitem{Zee1} V. Silveira and A. Zee, Phys. Lett. {\bf B161},
136 (1985).

\bibitem{Zee2}
D.E. Holz and A. Zee, Phys. Lett. {\bf B517}, 239 (2001) [arXiv:
hep-ph/0105284].

\bibitem{McD}
J. McDonald, Phys. Rev. {\bf D50}, 3637 (1994).

\bibitem{bento}
O. Bertolami, F.M. Nunes, Phys. Lett. {\bf B452}, 108 (1999)
[arXiv: hep-ph/9902439]; M.C. Bento, O. Bertolami, R. Rosenfeld,
L. Teodoro, Phys. Rev. {\bf D62}, 041302 (2000) [arXiv:
astro-ph/0003350]; M.C. Bento, O. Bertolami, R. Rosenfeld, Phys.
Lett. {\bf B518}, 276 (2001) [arXiv: hep-ph/0103340].

\bibitem{Posp1}
C.P. Burgess, M. Pospelov and T. ter Veldhuis, Nucl. Phys. {\bf
B619}, 709 (2001) [arXiv: hep-ph/0011335].

\bibitem{Posp}
C. Bird, P. Jackson, R. Kowalewski and M. Pospelov, Phys. Rev.
Lett. {\bf 93}, 201803 (2004) [arXiv:hep-ph/0401195]; C. Bird, R.
Kowalewski and M. Pospelov, Mod. Phys. Lett. {\bf A21}, 457 (2006)
[arXiv: hep-ph/0601090].

\bibitem{Zhu}
S.H. Zhu, [arXiv:hep-ph/0601224].

\bibitem{turner}
E.W.Kolb and M.Turner, {\it The Early Universe,  Addisson-Wesley,
1990}.

\bibitem{EXP}
M. Beneke, I. Efthymipopulos, M. L. Mangano, J. Womersley
(conveners) {\it et al.}, report in the {\it Workshop on Standard
Model Physics (and more) at the LHC}, Geneva, [arXiv:
hep-ph/0003033]
;E. Glover et al., hep-ph/0410110;
;T. Abe et al., {\it Linear Collider Physics Resource Book for
Snowmass 2001.} [arXiv: hep-ex/0106055, 0106056, 0106057].

\bibitem{SMhiggs}
A.~Djouadi,
[arXiv:hep-ph/0503172].

\bibitem{tchLHC}
J.A. Aguilar-Saavedra and G.C. Branco, Phys. Lett. {\bf B495}, 347
(2000); J.A. Aguilar-Saavedra, Phys. Lett. {\bf B502}, 115 (2001);
F. Larios, R. Martinez and M.A. PerezInt, J. Mod. Phys. {\bf A21},
3473 (2006) [arXiv: hep-ph/0605003].

\bibitem{hegunion} J. Gunion, B. Grzadkowski, X.-G. He,
    Phys. Rev. Lett. {\bf 77}, 5172(1996); J. Gunion and X.-G. He,
    Phys. Rev. Lett. {\bf 76}, 4468-4471(1996).


\bibitem{neutrino}
M. Frank and I. Turan, Phys. Rev. {\bf D74}, 073014 (2006) [arXiv:
hep-ph/0609069].

\bibitem{Soni}
G.~Eilam, B.~Haeri and A.~Soni,
Phys. Rev. {\bf D41}, 875 (1990).

\bibitem{2hdm} For reviews see, J. Gunion, H. Haber, G, Kane and S. Dawson,
{\it The Higgs Hunter's Guide}, Addison-Wesley, 1990; S. Bejar,
[arXiv: hep-ph/0606138]; R.A. Diaz, [arXiv: hep-ph/0212237].

\bibitem{jaas}
T.P. Cheng and M. Sher, Phys. Rev. {\bf D35}, 3484 (1987).

\end{thebibliography}
\end{document}